\documentclass[twocolumn,floatfix,superscriptaddress,longbibliography,nofootinbib]{revtex4-1}
\RequirePackage[sort&compress]{natbib}

\usepackage{multirow}
\usepackage{color}
\usepackage{soul}
\usepackage{tabularx}
\usepackage{amssymb,amsfonts,amsmath}
\usepackage{bm}
\usepackage[pdftex]{graphicx}
\usepackage{xcolor}
\usepackage{comment}
\usepackage{float}

\usepackage{supertabular}
\usepackage{longtable}
\usepackage[colorlinks = true,
            linkcolor = blue,
            urlcolor  = blue,
            citecolor = blue,
            anchorcolor = blue]{hyperref}

\usepackage{dcolumn}% Align table columns on decimal point
\usepackage{bm}% bold math
\usepackage[caption=false]{subfig}
\usepackage{xcolor}

\renewcommand{\(}{\left(}
\renewcommand{\)}{\right)}

\newcommand{\tr}[1]{\text{Tr}\(#1\)}
\renewcommand{\(}{\left(}
\renewcommand{\)}{\right)}

\newcommand{\norm}[1]{\left\lVert#1\right\rVert}

\def\@cite#1#2{$^{\mbox{\scriptsize #1\if@tempswa , #2\fi}}$}
\usepackage{xcolor}
\definecolor{RoyalBlue}{HTML}{4169e1}
\definecolor{black}{rgb}{0.0, 0.0, 0.0}
\definecolor{ForestGreen}{HTML}{228b22}
\definecolor{DarkGreen}{HTML}{006400}
\usepackage[normalem]{ulem} % added by Carlo!

\usepackage{setspace}
\usepackage{comment}
\usepackage{float}
\usepackage{todonotes}

\newcommand{\arsham}[1]{{\color{black} #1}}

\newcommand{\gb}[1]{{\color{black}#1}}
\begin{document}

\newcommand{\rev}[1]{{\color{black} #1}}

\preprint{APS/123-QED}

\title{Dismantling the information flow in complex interconnected systems}% Force line breaks with \\

\author{Arsham Ghavasieh}
\email[Corresponding author:~]{aghavasieh@fbk.eu}
\affiliation{Fondazione Bruno Kessler, Via Sommarive 18, 38123 Povo (TN), Italy}
\affiliation{Department of Physics, University of Trento, Via Sommarive 14, 38123 Povo (TN), Italy}

\author{Giulia Bertagnolli}
\affiliation{Fondazione Bruno Kessler, Via Sommarive 18, 38123 Povo (TN), Italy}
\affiliation{Department of Mathematics, University of Trento, Via Sommarive 14, 38123 Povo (TN), Italy}

\author{Manlio De Domenico}
\email[Corresponding author:~]{manlio.dedomenico@unipd.it}%
\affiliation{Fondazione Bruno Kessler, Via Sommarive 18, 38123 Povo (TN), Italy}
\affiliation{Department of Physics and Astronomy ``Galileo Galilei'', University of Padova, Padova, Italy}

\date{\today}% It is always \today, today,

\begin{abstract}

\rev{Microscopic structural damage, such as lesions in neural systems or disruptions in urban transportation networks, can impair the dynamics crucial for systems' functionality, such as electrochemical signals or human flows, or any other type of information exchange, respectively, at larger topological scales. Damage is usually modeled by progressive removal of components or connections and, consequently, systems' robustness is assessed in terms of how fast their structure fragments into disconnected sub-systems. Yet, this approach fails to capture how damage hinders the propagation of information across scales, since system function can be degraded even in absence of fragmentation ---e.g., pathological yet structurally integrated human brain.
Here, we probe the response to damage of dynamical processes on the top of complex networks, to study how such an information flow is affected. We find that removal of nodes central for network connectivity might have insignificant effects, challenging the traditional assumption that structural metrics alone are sufficient to gain insights about how complex systems operate. Using a damaging protocol explicitly accounting for flow dynamics, we analyze synthetic and empirical systems, from biological to infrastructural ones, and show that it is possible to drive the system towards functional fragmentation before full structural disintegration.}

\end{abstract}
            
\maketitle

\section{Introduction}

\begin{figure*}
    \centering
    \includegraphics[width=\textwidth]{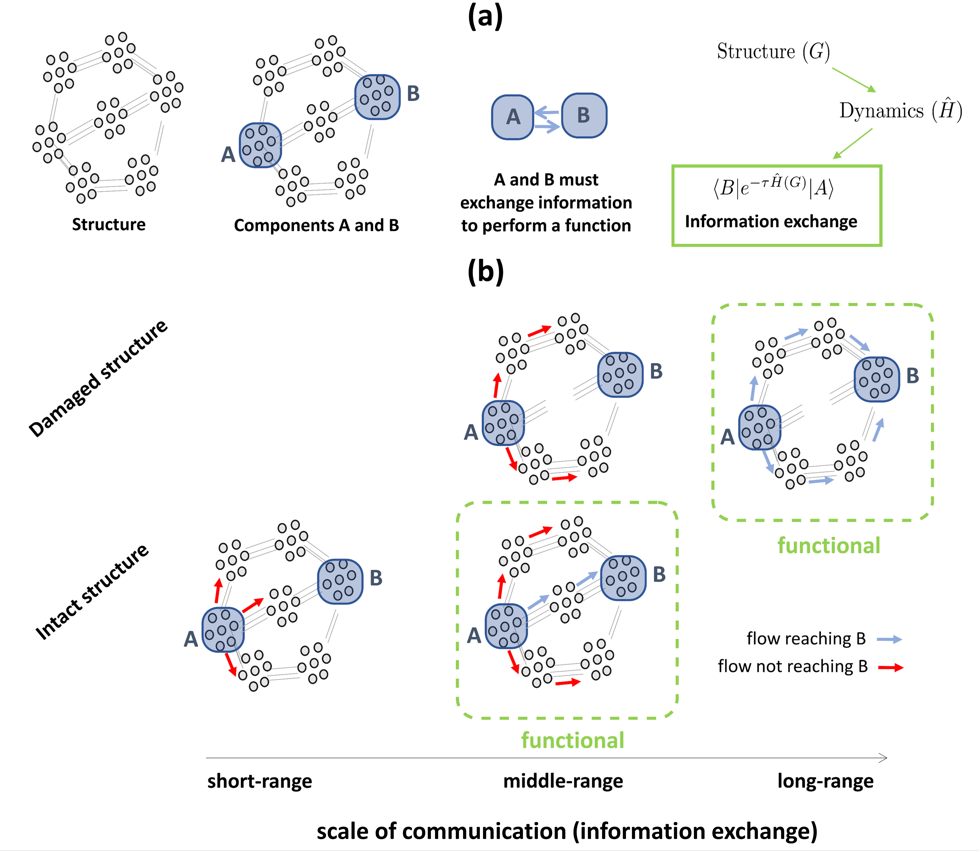}\rev{
    \caption{\textbf{Impairment of information flow and functional degradation.} (a) The structure of a complex system is shown, where two components A and B are highlighted in blue. To perform a system specific function, A and B must communicate---e.g., Broca and Wernicke areas in human brain for speech production, London and Berlin airports for flow of business people, industry section and academia exchanging ideas and knowledge for technological development--- and the flow of information between them can be modeled using differential equations coupling the structure, noted as $G$, and dynamical processes, governed by the control operator $\hat{H}$. (b) The dependence of functionality on the scale of information propagation in the system, from short- to long-range, and damage destroying part of the system. A and B can not perform the system specific function, if the communication range is short or in cases where the structure is damaged while the communication scale is not large enough. Note that the damage depicted here is not severe enough to dismantle the structure, yet it has a considerable effect on the flow and, thus, functionality.
    }
    }
    \label{fig:dismantling_vs_damage}
\end{figure*}

Despite the diversity of physical attributes, all complex systems can be viewed as large collections of units exchanging information to function properly\rev{---e.g., spreading of diseases through social systems, human flows between urban areas through transportation networks, financial transactions between financial agents in stock markets and electrochemical signals exchanged among neurons in the human brain}. Of course, regulation and maintenance of such pairwise communications necessitates the presence of an underlying structure, a network, exhibiting high resistance to disintegration~\cite{albert2000error,callaway2000network,holme2002attack,centola2008failure,trajanovski2013robustness,Allard, Jesus_Gomez,Radicchi} even under sever damage---e.g., genetic mutations in gene-gene interaction networks~\cite{manke2006entropic}, extinction of species in ecosystems~\cite{dunne2002network}, failure of Internet routers~\cite{doyle2005robust} or unavailability of transportation means~\cite{de2014navigability}. 
During the last decade, network integrity---i.e., the existence of \arsham{pathways between every pair of units} guaranteeing \arsham{that they can exchange information}---has been widely used as a proxy of network robustness, under progressive removal of units or connections~\cite{Iyer2013,de2014navigability}, indicating that in many empirical systems the full disintegration happens only if a large fraction of the system is damaged. 
While structural integrity is a necessary condition for information exchange, it is barely sufficient to quantify the effect of damage on the flow of information within the systems (See Fig.~\ref{fig:dismantling_vs_damage})\rev{, a property crucial for system's functionality. For instance, it has been shown that by analyzing the perturbation of information flow, one can distinguish the healthy brain from the pathological one, even in absence of significant structural differences among the two types of connectomes~\cite{Barbara_brain_2021}.} Accordingly, a number of methods have been developed to capture unit-unit communications, considering the coupling between the structure and dynamical processes governing the flow of information~\cite{Masuda2017,arenas2008synchronization,dedomenico2016physics} and to account for the heterogeneity and intervening of temporal and spatial information propagation scales~\cite{Arenas_synch,lambiotte_markov}. \rev{Yet, the robustness literature is still predominantly limited to structural analyses, missing a functional perspective.}

Of course, in special cases, like for very sparse networks, or when only short-range interactions are under investigation, the role played by the structure is expected to be dominant. In those cases, structural indicators provide a reliable description of the flow dynamics and shortest paths, sequences of minimum number of links connecting pairs of nodes, might reliably describe the flow pathways in the system.  
However, in most scenarios, \rev{damage can locally or globally perturb the information dynamics in empirical systems--- e.g., lesions impairing the flow of electrochemical signals within the larger or smaller sections of the brain, to failures of transportation systems hindering human flows at the level of districts, cities or regions---, and} ignoring the dynamics of how information spreads and the multiscale nature of interactions can lead to poor results. To this end, a novel framework is needed to go beyond the traditional structural paradigm and investigate functional robustness of interconnected systems, characterized in terms of the system's ability to maintain the flow exchange under random or targeted disruptions.

Here, we assess functional robustness in terms of the effect of unit removal on the information dynamics: a decrease in the average received information (ARI) per unit, modeled using diffusion processes coupled with the network, and an increase in the dispersion of units in the \rev{diffusion} manifold \rev{induced by the information dynamics}, quantified as their average squared diffusion distance (ASDD). 
We use statistical physics of complex information dynamics~\cite{sft2020} to identify and target the critical components of the system based on the impact of their removal on the Von Neumann entropy~\cite{entanglement2021}. We demonstrate that this method directly reduces the overlap between flows originating from different parts of the system and, consequently, leads to efficient dismantling of the information flow\arsham{---i.e., functional dismantling}. 

Our results show that while attack strategies based on topological indicators such as degree, closeness and betweenness are effective in a limited number of scenarios, they are not distinguishable from random failures when the network is not sufficiently sparse or when the mid- or long-range communications are the target. \rev{Instead, our approach can identify the most central components of the human connectome, the neural network of the C. Elegans, European airline network and Chilean power grid, at multiple propagation time scales, whose removal effectively disturbs the flow of information in those systems.}

\section{From structural to functional robustness} 

The size of the largest connected component (LCC)---i.e., the number of nodes in the largest group of nodes that are connected together via sequences of links---has been extensively used as a proxy for structural robustness of networks~\cite{albert2000error,holme2002attack,Iyer2013}. In this case, the network is considered to be structurally dismantled when the size of LCC drops, consequent to removal of a fraction of links or nodes representing the damage imposed on the system, either in the form of random failures or targeted attacks. While the former can be easily simulated, using an algorithm that randomly removes nodes from the network and keeps a record of the size of LCC, there is no unique way to model the latter. To compensate, a variety of centrality measures have been introduced to rank the nodes according to different criteria of importance, and remove them, one by one, aiming to pose the maximum possible damage to the structure~\cite{Havlin_2018}. However, we still lack a universal centrality measure proven to be the most effective in quick structural dismantling of all network types.

In general, there are two types of attack strategies: i) static, where the ranking of the nodes is computed only once at the beginning of the algorithm and ii) iterative, where the centrality of nodes are updated, after every node removal. The first category has significantly lower computational cost, while the second one is more effective in leading networks to quick structural impairment. On the one hand, iterative betweenness ---i.e., a centrality that ranks the nodes according to the number of shortest paths crossing them---, shows impressive performance~\cite{Havlin_2018} with reasonably low computational complexity. On the other hand, new measures of centrality have been introduced to combine topological metrics with meta data analysis~\cite{feature_based_2021} or based on machine learning~\cite{ML_2021}, geometry~\cite{geometry_centrality_2019} and, very recently, statistical physics~\cite{entanglement2021}.

In this work, we aim to study functional robustness, in contrast with the structural. Our goal is to test the performance of the widely used, yet purely topological, metrics including iterative degree, betweenness, eigenvector and closeness centrality, in dismantling the flow of information, instead of LCC. Additionally, we compare the structural metrics with a recent centrality measure named entanglement, that ranks the nodes according to the impact of their removal on the diversity of flow pathways in the system, providing a suitable candidate for effective functional dismantling. It is worth mentioning that entanglement centrality has also been previously studied against a large number of widely used structural and dynamical centrality measures and shown to outperform them, or perform as well as the best of them, in dismantling a wide range of networks structurally~\cite{entanglement2021}. For this reason, here, we study the effectiveness of the best performing structural measures in comparison with entanglement, excluding other dynamical metrics. 

In the following, we firstly introduce ARI and ASDD, as proxies for functional robustness. Then, we review the definition of entanglement centrality derived from statistical physics of complex information dynamics. Finally, we show that topological metrics fail to identify the nodes central for the flow dynamics, under a number of scenarios. We demonstrate that, surprisingly, the effectiveness of attacks guided by topological measures on the functional robustness is not distinguishable from random damage. This result highlights that topological information, including the distribution of degree and shortest paths, is not sufficient to capture the information flow in synthetic and empirical networks and, therefore, shortest paths are not determinant of the node-node interactions.

\section{Quantifying functional robustness}

\begin{figure*}
\centering
\includegraphics[width=\textwidth]{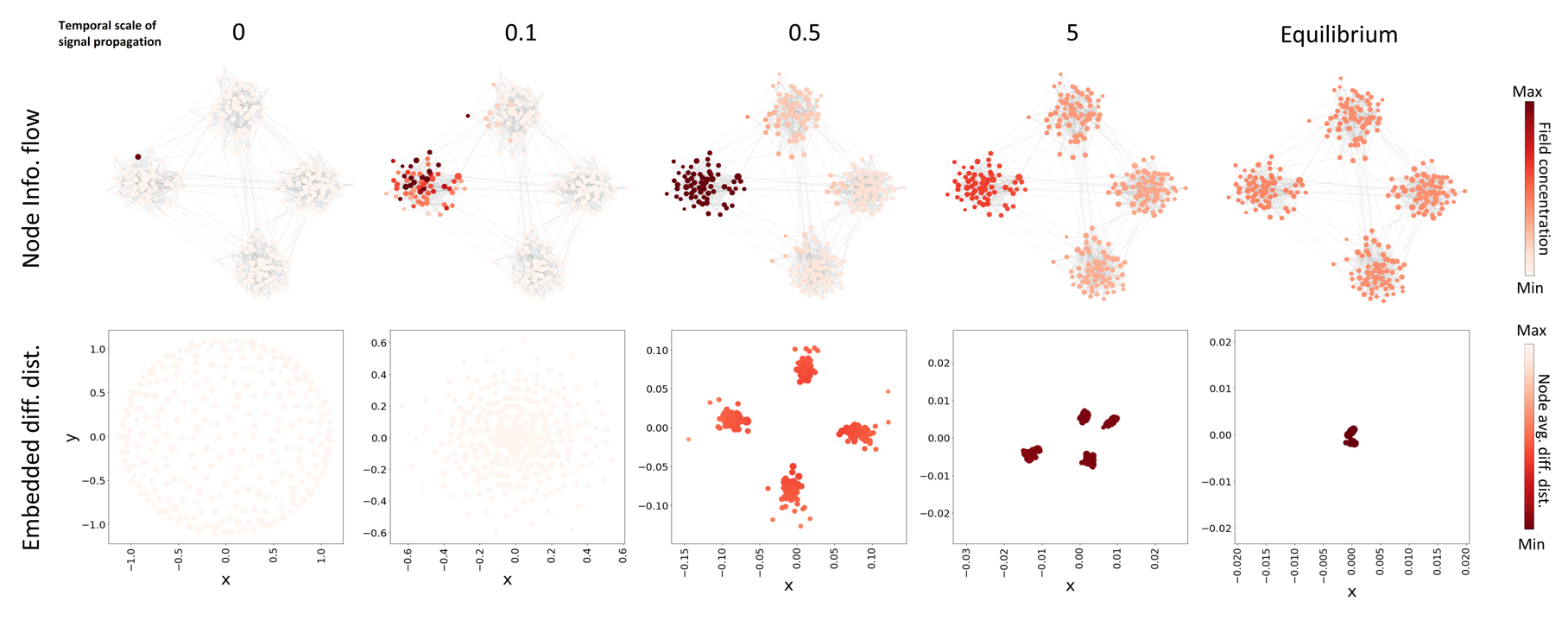}
\caption{\label{fig:communcation}\textbf{Multiscale information propagation.} Top row shows the propagation from an arbitrary node (having high field concentration at $\tau = 0$). At small temporal scales, the field propagation is limited to the locality ($\tau = 0.1,0.5$), while the global interactions are allowed at larger temporal scales ($\tau=5$). The equilibrium of the dynamics happens when the field reaches the final distribution, independent of its initial conditions. Below each plot in the top row, the propagation geometry constructed from~\eqref{eq:diffu-dist}, embedded in 2 dimensions using multidimensional scaling, is shown. Each dot corresponds to a node and node colors indicate the average diffusion distance between all vertices in the network. The nodes become closer in the diffusion manifold as one tunes $\tau$ to larger values, from separate units to functional modules and, eventually, reaching full integration. Similar coloring can be obtained using average squared diffusion distance instead of average diffusion distance. }
\end{figure*}

Different levels of abstraction can be implemented in modeling a complex phenomenon, from agent based modeling~\cite{ABM}, where one tries to insert every available detail into the equations, to the mean-field approaches, where even the agents might be coarse grained~\cite{Coarse_grained,mean_field} into fields with no regard for the connectivity patterns, heterogeneity, etc. Comparing the two approaches, one immediately observes a trade-off. High resolution modeling is often preferred when exact calculations are necessary for decision or policy making. Yet, low resolution modeling is also powerful, as it captures the most important factors and their associations, providing understanding at the expense of precision. For instance, most network science studies during the last two decades ground themselves in structural analysis of real-world systems, counting the number of connections, finding the geodesic distances as plausible transportation routes and studying the modularity and hierarchy of the structure, all based on distribution of links among the nodes. One abstraction level above the structural analysis is to couple the network with dynamical processes, such as classes of diffusion, that are general enough to describe a wide range of transport phenomena~\cite{transport2020}, at least to the first order approximation, and suitable to model communications between the nodes. 

To mathematically describe the coupling, we identify the nodes of a network as canonical vectors $|x_{i}\rangle, \, i = 1, 2, \dots, N$ and encode their connections in the operator $\hat{W}$ which in the space of nodes-- shaped by the canonical vectors $|x_{i}\rangle$-- represents the adjacency matrix, where $\langle x_{j}| \hat{W} |x_{i}\rangle = W_{ij}$ is the weight of the link from $i$-th to $j$-th node, \rev{in accordance with notation of a fundamental reference of this work~\cite{sft2020}. The field $| \phi(\tau)\rangle$ is assumed on top of the network, and its amount at node $i$ at time $\tau$ is shown as $\langle x_{i}|{\phi}(\tau)\rangle$. The flow of the field from one node to another will be used as a proxy for information exchange between the two}. The evolution is governed by the linear or, in case of non-linear dynamics, linearized equation 

\begin{eqnarray}\label{eq:master}
\partial_{\tau}|\phi(\tau)\rangle= -\hat{H}(\hat{W}) |\phi(\tau)\rangle ,
\end{eqnarray}
where $\hat{H}(\hat{W})$ is a control operator. Note that the equation is exact for a range of dynamical processes including random walks, consensus dynamics, synchronization near the meta-stable manifold and continuous diffusion. As explained before, in this work we focus on diffusion dynamics where the control operator becomes the Laplacian matrix $\hat{H}(\hat{W})=\hat{K}(\hat{W})-\hat{W}$, with $\hat{K}(\hat{W})$ being the diagonal degree matrix where $\langle x_{j}|\hat{K}(\hat{W}) |x_{i}\rangle = \delta_{ij} k_{i}$ with the Kronecker delta function $\delta_{ij}$ that is equal to $1$ when $i=j$ and $0$ otherwise, and $k_{i} = \sum\limits_{j=1}^{N}W_{ij}$ being the degree of node $i$.

Solving the linear equation, one obtains the propagator $\hat{U}(\tau,\hat{W})=e^{-\tau\hat{H}(\hat{W})}$. Consequently, information flow from node $i$ to node $j$ can be described using a propagator method similar to the Euclidean path integrals $\langle x_{j}|\hat{U}(\tau, \hat{W})|x_{i}\rangle$, and by changing $\tau$ from low to high values, the propagator can encode short-, middle- and long-range communications~\cite{sft2020}. 

In this case, the average received flow per node reads

\begin{eqnarray}
\label{eq:ARI}
\mathcal{A}(\tau,\hat{W})=\frac{1}{N}\sum\limits_{i\neq j} \langle x_{j}|e^{-\tau \hat{H}(\hat{W}) }|x_{i}\rangle,
\end{eqnarray}
where $N$ is the number of nodes in the network. Significant damage is expected to lower the average received information per node. In the following sections, we apply random and targeted removal of the nodes to a range of synthetic and empirical networks and study the alterations of ARI along dismantling trajectories, keeping a record of the effect of progressive damage on the communications. Note that the nodes removed in the process do not exchange information with the network, having no impact on the summation in~\eqref{eq:ARI}, they can be assumed excluded from the analysis. Therefore, after removal of $m$ nodes, the denominator of~\eqref{eq:ARI} would be equal to $N-m$.

Based on the same propagator, one can consider the Euclidean distance between the propagation of the field at time $\tau$ from nodes $i$ and $j$
\begin{align}\label{eq:diffu-dist}
    D_{\tau}(i, j) & = \norm{ e^{-\tau \hat{H}(\hat{W}) } | x_i\rangle - e^{-\tau \hat{H}(\hat{W}) } | x_j\rangle }_2\\
    & = \sqrt{\sum_{k} \left(\langle x_k | e^{-\tau \hat{H}(\hat{W}) } | x_i \rangle - \langle x_k | e^{-\tau \hat{H}(\hat{W}) } | x_j \rangle\right)^2} \nonumber
\end{align} 
which induces a metric, called diffusion distance~\cite{dedomenico2017diffusion}, on the network \rev{and, consequently a (geo)metric structure on the set of nodes, called diffusion geometry~\cite{dedomenico2017diffusion}.}
According to $D_{\tau}$ nodes $i$ and $j$ are close if the flows emanating from them are similar or, in probabilistic terms, if the probability that two independent random walkers starting from nodes $i$ and $j$ respectively meet in any node $k$ at time $\tau$ is high.
\rev{Observe also that the information dynamics maps each node $i \in V$ to a point $e^{-\tau \hat{H}(\hat{W}) } | x_i\rangle$ on a hyperspace of $\mathbb{R}^N$~\cite{Bertagnolli2021}, called diffusion space, where nodes connected by many short walks lie close to each other.}
Since the diffusion distance integrates the information on the connectivity of the network at different scales, it is also sensitive to those structural changes forming bottlenecks in the information flow.

We summarize the geometric information provided by the diffusion distance through the average squared diffusion distance (ASDD)
\begin{equation}\label{eq:ASDD}
    \mathcal{D}^2(\tau, \hat{W}) = \frac{1}{2 N^2} \sum_{i, j} D_{\tau}^2 (i, j)
\end{equation}
which can be shown to be a measure of dispersion in the diffusion space, exploiting the fact that the propagator $e^{-\tau \hat{H}(\hat{W})}$ is a stochastic matrix---the Laplacian $\hat{H}(\hat{W})$ is, indeed, a $Q-$matrix~\cite{Norris1997,Bertagnolli2021} and, under particular assumptions on $\phi$~\eqref{eq:master} describes an edge-centric random walk~\cite{Masuda2017}, see the Appendix~\ref{appx:diff-geom}.
Consequently, we expect that the network dismantling increases the dispersion of the nodes (points) in the diffusion space. 

\arsham{For a visual illustration of information propagation and the functional geometry of networks, see Fig.~\ref{fig:communcation}. Obviously, when $\tau$ is sufficiently small, information exchange is limited to the locality of the nodes, among first neighbors, and through the shortest paths connecting them to the spatially distant ones. \gb{Also, at this temporal scale, the diffusion distance between the nodes is expected to be strongly determined by the shortest paths connecting them---i.e., the local geometry of the network.} While, as one increases $\tau$ in a network that is not extremely sparse, the longer paths \gb{(and their number)} become important, modulating the amount of exchanged field among the nodes and the mutual diffusion distance between them. 
In this regime, we expect that both ARI and ASDD can not be understood in terms of purely structural metrics such as betweenness and degree. } 

\section{Entanglement centrality}

Originally, the notion of density matrices has been introduced in quantum mechanics to capture the pairwise coherence between quantum states in a physical system~\cite{Fano_1957}. Similarly, complex networks, as collections of pairwise connections between objects, cannot be fully described by vectors or distribution functions, without information loss. Therefore, following this analogy and inspired by the framework of quantum statistical physics, the state of complex networks has been derived in terms of density matrices~\cite{sft2020}, describing statistical ensembles of stream operators given by the outer product of eigenvectors of a control operator $\hat{H}$ that guides the flow dynamics--- note that such density matrix has been only used to analyze classical complex networks and its applicability in case of quantum complex networks is still open to be explored. The framework has been successfully applied to analyze and improve the transportation properties of social and infrastructural multiplex networks~\cite{transport2020}, cluster the human microbiome~\cite{de2016spectral} extract the mesoscale organization and functional diversity of the human brain~\cite{Nicolini_2020,Barbara_brain_2021}, and characterize functional modules in fungal networks~\cite{Fungal_2021}. 

Interestingly, the Von Neumann entropy has been shown to be a measure of diversity of flow pathways in the systems~\cite{sft2020}. Consequently, the effect of node removal on the diversity of flow pathways, reflected in the Von Neumann entropy, has been used as a measure of centrality, named \emph{network entanglement}. The effectiveness of network entanglement in structural dismantling has been compared against a range of mostly used structural and dynamical centrality measures~\cite{entanglement2021}. It has been shown that attacks guided by entanglement dismantled synthetic and empirical networks faster than other measures, or as fast as the best of them. Yet, the computational complexity of entanglement has been shown to be higher than many of these measures, being around $\mathcal{O}(N^{3})$ after mathematical approximations and treatment. This makes other structural and dynamical metrics, even though with lower performance in guiding structural or functional attacks, more scalable. Nevertheless, entanglement allows for interpretation and better understanding of the multiscale nature of node importance in terms of the effect of its removal on the diversity of flow pathways and, therefore, assessing functional robustness, as shown in the following.
For a more comprehensive review of network entanglement and the slightly different version of it used here, please see appendinx.~\ref{appx:entanglement}.

In the following, we compare the effectiveness of attack strategies guided by entanglement with those based on structural metrics including the ones with highest performance in structural dismantling, like iterative betweennes.

\section{Synthetic networks} 

\begin{figure*}[t]
    \centering
    \includegraphics[width=.9\textwidth]{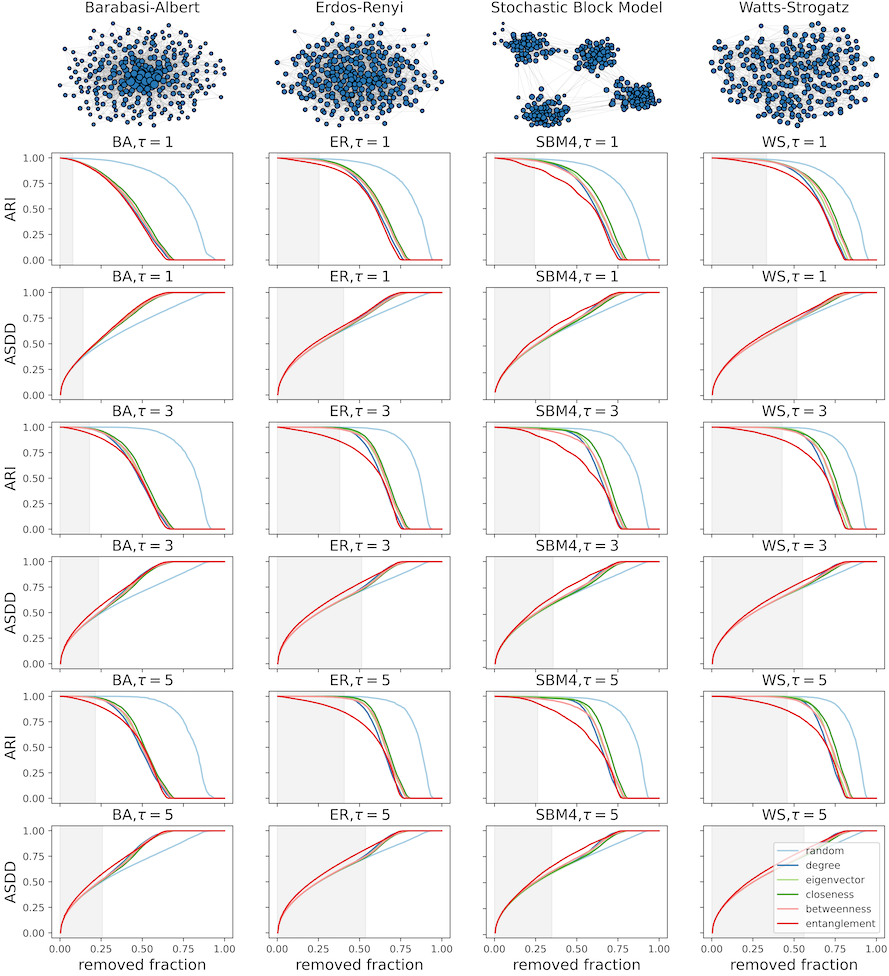}
   \caption{\label{fig:Toy}\textbf{Functional dismantling of synthetic networks.}
   10 independent realization of 4 classes of network ($N= 256$ and average degree $\approx 12$) have been considered: Barabasi-Albert (BA), Erods-Renyi (ER), Stochastic Block Model with four communities (SBM4) and Watts-Strogatz (WS). 
   ARI and ASDD at different temporal scales ($\tau = 1,3,5$) have been studied under random failures and targeted attacks based on structural centrality measures including iterative degree, betweenness, closeness and eigenvector and entanglement, as a functional metric. The gray area in each plot shows the region in which the effectiveness of structural measures is comparable to random selection of nodes. Under all scenarios considered here, entanglement centrality is able to impose larger impact on ARI and ASDD.}
\end{figure*}

We study the functional robustness of four different network classes, including Barabasi-Albert~\cite{BA_1999}, Erods-Renyi~\cite{ER}, stochastic block model with four communities and Watts-Strogatz~\cite{Watts1998} models, as useful models of real-world systems. For each class, 10 independent realizations with $N= 256$ nodes and average degree approximately equal to $12$ are evaluated under random failures and targeted attacks guided by structural measures including iterative degree, betweenness, closeness and eigenvector centrality. 

In most cases and as long as the size of damage---i.e., the percentage of removed nodes---is not considerably large, the effectiveness of these structural metrics on ARI and ASDD is not distinguishable from that of random failures (See Fig.\ref{fig:Toy}). It is worth highlighting that even the performance of iterative betweenness, which is known to be highly efficient in dismantling the networks structurally, is comparable to random disruptions unless the size of damage is considerably large. 
The only counter example among the considered cases is for Barabasi-Albert networks at small temporal scales, $\tau=1$\gb{, due to the characteristic heterogeneity of its degree distribution.} 
Generally, it is evident that the ability of structural measures in capturing centrality is higher in case of smaller propagation times. 
\gb{This means that if the flow dynamics on a real network is local---e.g., the spreading of a niche content on an online social network like Twitter cannot happen forever and reach everyone since after some time $T>0$ the tweet is scrolled down the timeline---then attacking topologically central nodes is efficient also for dismantling the information flow.}
Of course, this is expected as small $\tau$ characterizes local interactions between the nodes and the local interactions can be directly described by the adjacency matrix, regardless of the coupling between network and diffusion dynamics.

According to our results, entanglement as a functional metric outperforms the structural ones, under almost all considered circumstances. This is even more interesting in the light of the previous works, showing that this measure has no particular correlation with and cannot be captured by any of a wide range of structural and dynamical centrality measures~\cite{entanglement2021}.

\section{Empirical networks}

\begin{figure*}[t]
    \centering
    \includegraphics[width=.9\textwidth]{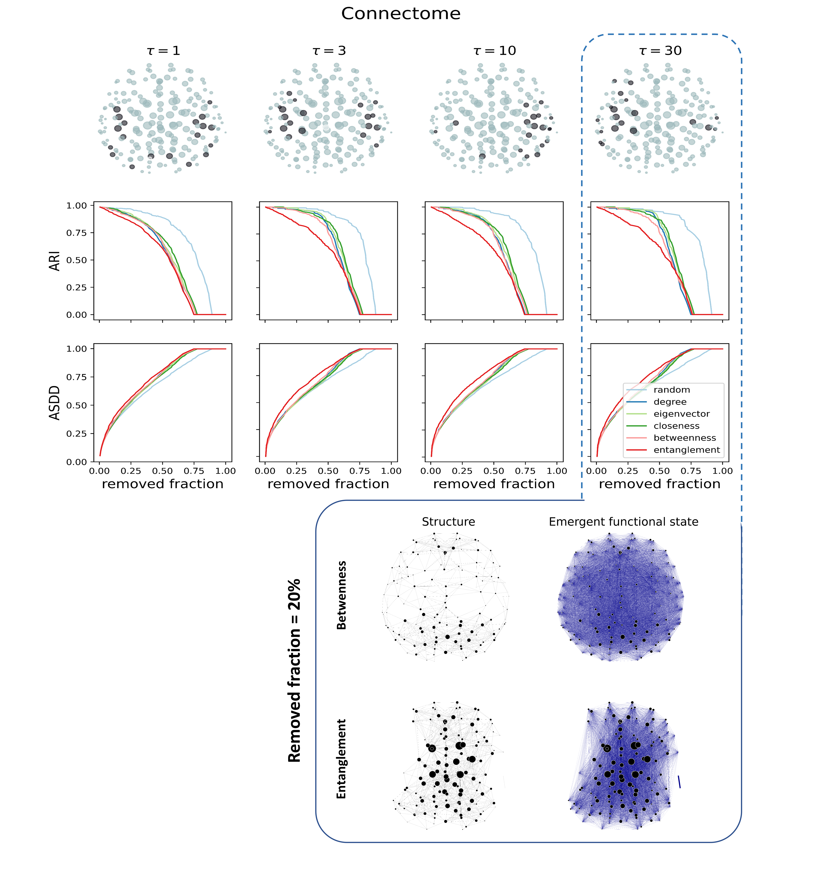}
    \caption{\label{fig:brain}\textbf{Functional dismantling of the Human Connectome.}
    Attacks based on entanglement centrality provide the upper bound for functional damage on the human connectome, among the considered centrality measures. The overall effectiveness of structural measures are indistinguishable from each other and comparable with random failures at large temporal scales. The spatial locations of nodes match the positions of brain regions given by the dataset. The size of the nodes is proportional to their degree. Dark gray indicates the top 10\% of the nodes according to entanglement centrality, at specific $\tau$. As an example, the effect of removal of 20\% of nodes guided by betweenness and entanglement centrality, on the structure and \emph{emergent functional state}~\cite{Fungal_2021}--- i.e., a network in which the edge weights are given by the amount of flow exchange between the nodes--- has been illustrated, at $\tau=30$.
    }
\end{figure*}

\begin{figure*}[t]
    \centering
    \includegraphics[width=\textwidth]{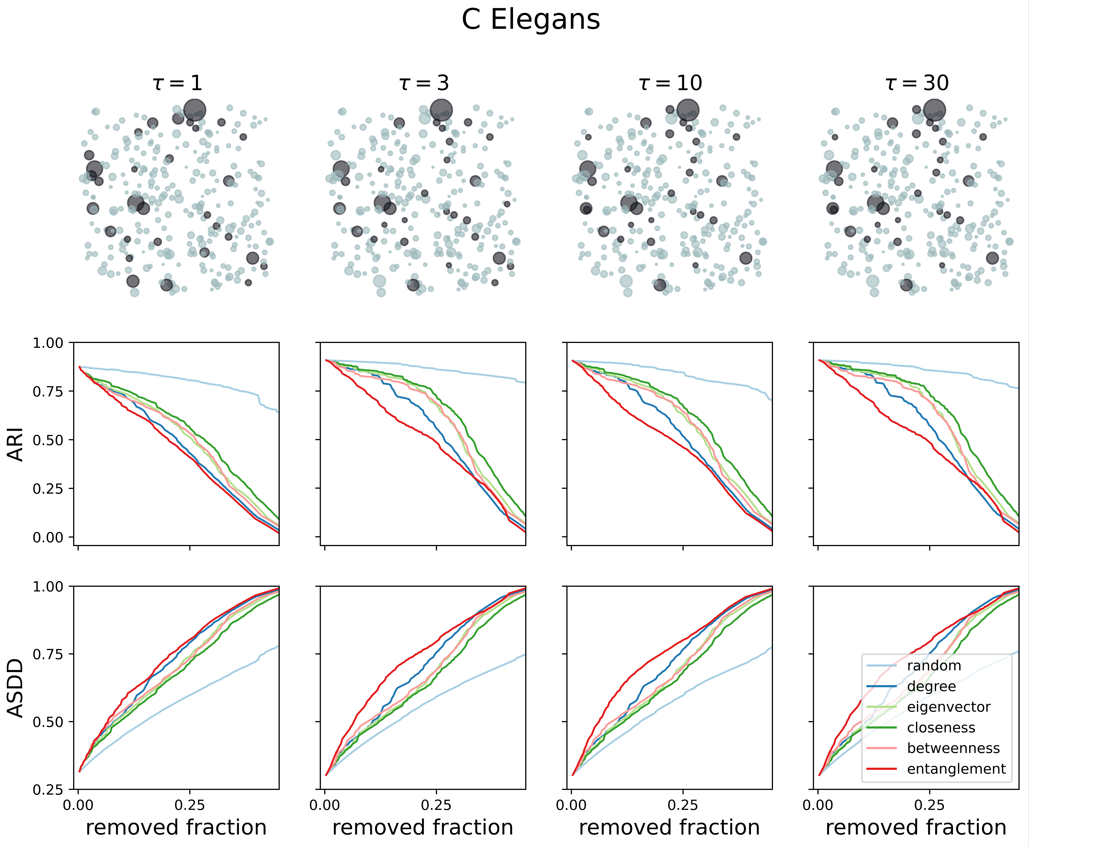}
    \caption{\label{fig:CElegans}\textbf{Functional dismantling of C. Elegans neural system.} The neural network of the nematode worm \textit{C. Elegans} ($N=297$) is considered. The effect of random failures and targeted attacks guided by iterative betweenness, degree, eigenvector, closeness and entanglement centrality measure on ARI and ASDD is plotted at multiple propagation time scales $\tau=1,3,10,30$. Attacks based on entanglement centrality provide the upper bound for damage among the considered centrality measures. The spatial locations of nodes match the positions of brain regions given by the dataset. The size of the nodes is proportional to their degree. Dark gray indicates the top 10\% of the nodes according to entanglement centrality, at specific $\tau$. }
\end{figure*}

\begin{figure*}[t]
    \centering
    \includegraphics[width=\textwidth]{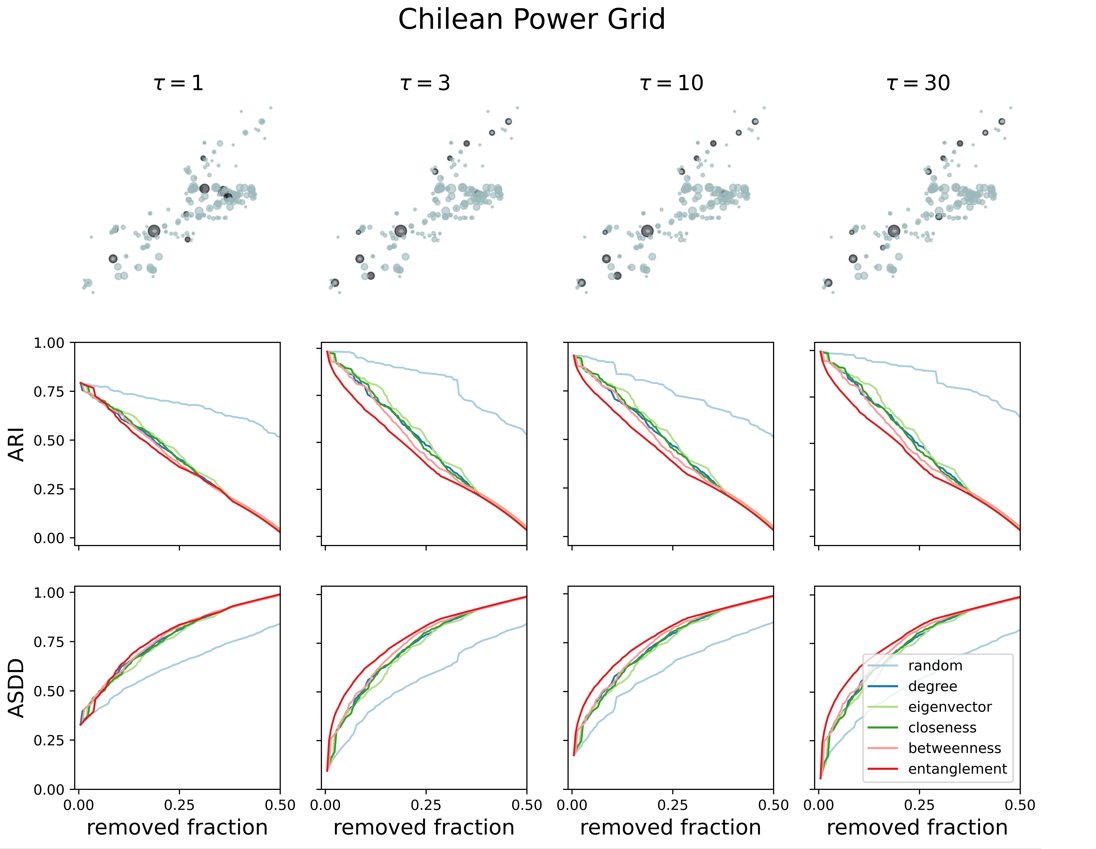}
    \caption{\label{fig:grid}\textbf{Functional dismantling of Chilean Power Grid.} The Chillean power grid network ($N=218$) indicating power plants and
    substations is considered. The effect of random failures and targeted attacks guided by iterative betweenness, degree, eigenvector, closeness and entanglement centrality measure on ARI and ASDD is plotted at multiple propagation time scales $\tau=1,3,10,30$. Attacks based on entanglement centrality provide the upper bound for damage among the considered centrality measures. The overall effectiveness of structural measures are almost indistinguishable from each other. The spatial locations of nodes match the positions of brain regions given by the dataset. The size of the nodes is proportional to their degree. Dark gray indicates the top 10\% of the nodes according to entanglement centrality, at specific $\tau$. }
\end{figure*}

\begin{figure*}[t]
    \centering
    \includegraphics[width=\textwidth]{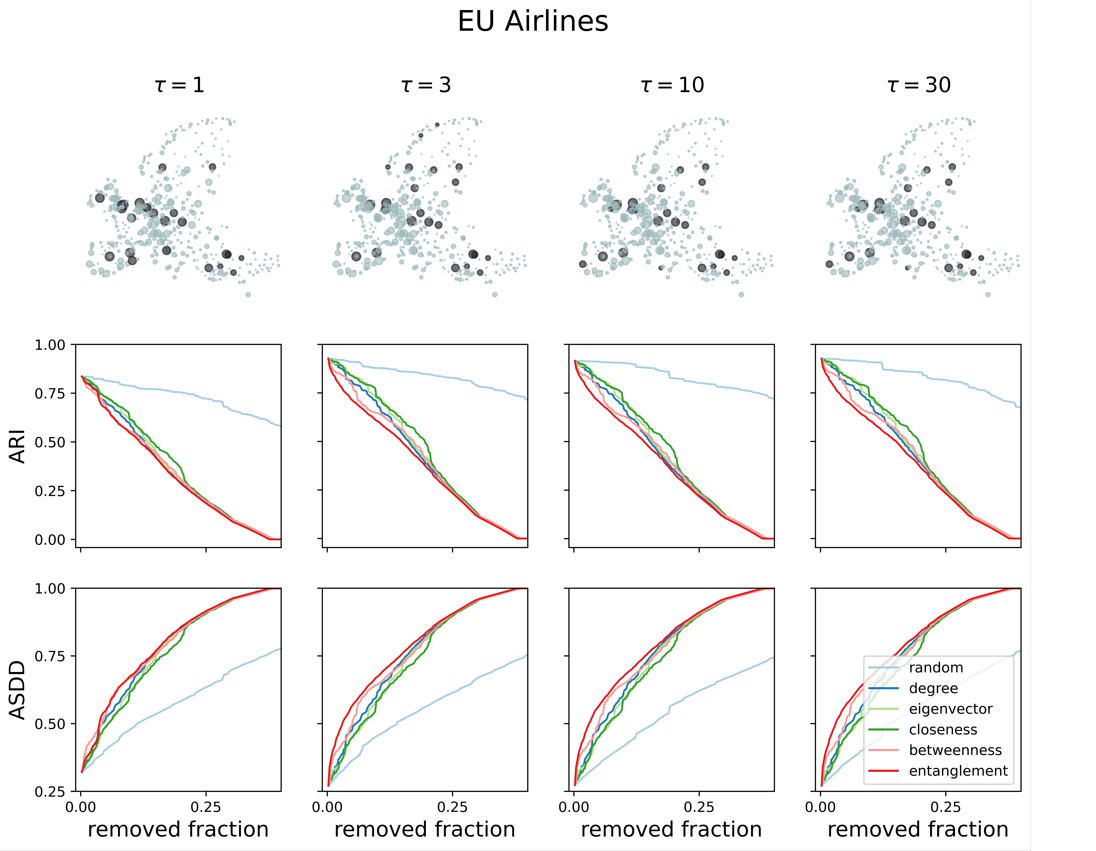}
    \caption{\label{fig:EUAir}\textbf{Functional dismantling of EU Airlines.} the European airlines network ($N=450$) is considered. The effect of random failures and targeted attacks guided by iterative betweenness, degree, eigenvector, closeness and entanglement centrality measure on ARI and ASDD is plotted at multiple propagation time scales $\tau=1,3,10,30$. Attacks based on entanglement centrality provide the upper bound for damage among the considered centrality measures. The overall effectiveness of structural measures are almost indistinguishable from each other. The spatial locations of nodes match the positions of brain regions given by the dataset. The size of the nodes is proportional to their degree. Dark gray indicates the top 10\% of the nodes according to entanglement centrality, at specific $\tau$.}
\end{figure*}

In this section, we analyze the functional robustness of four empirical interconnected systems, including two natural and two man-made: an averaged human connectome ($N=188$) provided by the NKI-Rockland sample including brain regions and their structural connections constructed using Diffusion Tensor Imaging (DTI)~\cite{Rockland_data}, the neural network of the nematode worm \textit{C. Elegans} ($N=297$) representing neurons linked by their neural junctions~\cite{Watts1998}, the reduced Chillean power grid network~\cite{Chilean_Power,Chilean_Power_data} ($N=218$) representing power plants and
substations, and the European airlines network ($N=450$)~\cite{European_airlines_data}, originally a multilayer with thirty-seven layers that we aggregate into a single layer network. 

While most well-known structural centrality measures have been generalized to be able to cope with weighted networks, their success is mostly shown for binary adjacency matrices in the literature. For this reason, these measures are expected to be less competent for the analysis weighted network against a functional measure like entanglement. Consequently, for the weighted real-world networks studied here, such comparison can be unfair and, more importantly, the result of such juxtapositions would not support the core message of this study, which is to show that structural information can not provide a multiscale proxy for node-node interactions. Therefore, we binarize the weighted adjacency matrices, keeping only the elements above the threshold of one sigma---i.e., one standard deviation above the average of elements---and use the binarized networks to perform the analysis of functional robustness.

According to our results, attacks guided by entanglement exhibit an upper bound for the damage reflected in ARI and ASDD, especially when long range interactions are under investigation, in almost all scenarios (See Figures~\ref{fig:brain},~\ref{fig:CElegans},~\ref{fig:EUAir} and ~\ref{fig:grid}). 
In the human connectome and C. Elegans network the gap between structural metrics and entanglement is more evident, compared with the technological networks. 
\gb{Brain networks are characterized by different interlinked regions, which are able to i) carry on specific tasks processing information inside each module and ii) integrate information functioning as a whole and enabling cognition, thanks to long-range links, a multiscale modular organization~\cite{Betzel2017}, rich-clubs and cores~\cite{Van2011rich} etc. 
Not only the network topology shapes flows in the brain, but also different communications dynamics do~\cite{Avena2018}, since the creation of each synaptic connection involves a trade-off between costs (in terms of energy) and benefits (e.g., increasing the routing or the diffusion efficiency of the network, or the network resilience). Hence, removing highly entangled nodes, may not disrupt the network topologically but may have a non-trivial impact the brain functions.}
The multiscale property of entanglement allows for the identification of different groups of nodes that are important for short, middle and long range signaling. For example, in the human connectome at $\tau=1$ and $\tau=3$ the important regions are distributed almost equally across the right and left hemispheres with a tilt from the occipital at $\tau=1$ to the temporal areas at $\tau=3$. However, at middle propagation scales, $\tau=10$, most central regions populate in the right hemisphere and at the large scales, $\tau=30$, this asymmetry shifts to the left hemisphere.   

As mentioned earlier, for very sparse networks or when one is interested in exploring only short range interactions limited to the locality, the structure is expected to plays a dominant role and, therefore, structural centrality measures become reliable for dismantling the flow. Our results clearly confirm the former expectation, as the structural centrality measures exhibit a better performance at small temporal scales, where nodes interact inside their neighborhood and the adjacency matrix provides a good proxy for the flow, in all considered cases. 
It can also be mathematically demonstrated in terms of a Taylor expansion of the propagator at $\tau \ll 1$, as $\hat{U}(\tau,\hat{W})\approx \hat{I}-\tau \hat{H}(\hat{W}) = \hat{I} - \tau \hat{K}(\hat{W}) + \tau \hat{W}$. The off-diagonal elements are directly given by the adjacency operator $\hat{W}$, with $\hat{I}$ being the identity and $\hat{K}(\hat{W})$ the diagonal degree matrix. The latter expectation is also supported by our empirical network analysis. As shown for European Airlines and Chilean Power Grid, the sparseness of technological networks make them more vulnerable to functional impairment under attacks based on structural metrics. For example, the Chilean power grid network with $N=218$ has only $527$ links. 
Nevertheless, even in such cases, entanglement centrality outperforms the structural measures, provides insights into the multiscale nature of node importance for information flow \gb{and of the process-driven network geometry}, and identifies groups of nodes that are specifically important at every scale \gb{and whose removal has also the strongest impact on the diffusion geometry of the network.}

\section{discussion}

Instead of the traditional approach to the problem of network robustness that is based on quantifying the resistance of networks to structural impairment, we explored the functional effect of progressive damages. To this aim, exploiting diffusion dynamics coupled with the structure, we introduced two descriptors: the average received information per node and the average squared diffusion distance between the nodes. Thus, we assessed the effect of random and targeted attacks  guided by a number of widely used centrality measures on these metrics, in a broad range of scenarios including synthetic and empirical networks.

Our results indicate that removing the nodes with high topological centrality, for example the ones having high degree or betweenness, \arsham{might have a} surprisingly insignificant effect on the flow dynamics. More specifically, the effectiveness of attack strategies guided by such structural metrics \arsham{can be indistinguishable} from that of random removal of nodes, most evident when mid- to long-range propagation time-scales are considered and the network is not extremely sparse. This finding stands as further evidence supporting the paradigm shift happening in today's network science, based on the fact that \arsham{structural information, on their own, can be unreliable}. For instance, the \arsham{limitation} of iterative betweenness in identifying the important nodes suggests that the transport phenomena can not be captured only in terms of shortest paths, an approximation taken for granted in the literature. 
Alternatively, we have shown that using statistical physics of complex information dynamics, one can identify nodes whose removal directly impacts the diversity of flow pathways in the system as a whole \gb{and the system's geometry induced by the information diffusion}, leading to a quick impairment of the flow dynamics \arsham{in a range of synthetic and a multitude of empirical systems across scales, from the neural network of C. Elegans and Human Connectome to Chilean Power Grid and European Airlines.}

Overall, apart from the practical aspects such as flow dismantling and multiscale identification of central units, our work provides insights into both applicability and limitation of structural metrics in capturing complex collective phenomena, such as information dynamics. 
In other words, the functional robustness framework has been able to differentiate two regimes: \arsham{i) where the role played by the network is dominant and structural metrics can be reliably used as fast and effective tools to proxy interactions and ii) where the structural measures are not sufficiently sophisticated to capture the complexity arising from the coupling between structure and dynamics.}
\gb{In i) the system's units are far apart in the diffusion space, because at very small time scales the flows are localized around their source and the dynamical proximity of the nodes in the network, captured by the diffusion distance, depends mostly on the local connectivity. While in regime ii) the intermediate time scales of the dynamics allow the integration of local and global connectivity information, so that the geometry is able to reveal not only the fine details of the topology, but also the interplay between the global structure and the flow pathways between pairs of nodes.
The link between geometry, flow dynamics and functional robustness is the spectrum of the Laplacian matrix: Its eigenvalues quantify the diversity of flow pathways in the system, but also the characteristic times of a random walk among its units and the shape of the system in its diffusion space.}

Our work introduces a novel functional perspective into the robustness analysis of interconnected systems, indicates the failure of structural metrics in identifying the nodes central
for information dynamics and \arsham{highlights the power of methods grounded in statistical physics and geometry in unraveling the complex interplay between the structure and dynamics.}

\bibliography{biblio}

\appendix

\section{Diffusion geometry}\label{appx:diff-geom}

\paragraph{On the diffusion distance}
We start by showing that the Laplacian matrix $\hat{H}(\hat{W})=\hat{K}(\hat{W})-\hat{W}$ is a $Q-$matrix on the set of nodes $V$ of the network $G= (V, E)$ or, more precisely, that $-\hat{H}(\hat{W})$ does.
In the jargon of probability theory, a $Q-$matrix~\cite{Norris1997} on $V$ is a matrix $Q=\left(q_{ij} : i, j \in V\right)$ such that
\begin{itemize}
    \item[(i)] $0 \leq -q_{ii} < +\infty$ for all $i$
    \item[(ii)] $q_{ij} \geq 0$ for all $i \neq j$
    \item[(iii)] $\sum\limits_{j \in V} q_{ij} = 0$ for all $i$.
\end{itemize}
The check is immediate.
This is equivalent~\cite[Thm. 2.1.2]{Norris1997} to proving that $e^{-\tau \hat{H}(\hat{W})}$ is a stochastic matrix for all $\tau \geq 0$, i.e., it has non-negative elements and its rows sum up to 1.
From a $Q-$matrix $Q$ it is always possible to obtain its jump matrix $\Pi = \left(\pi_{ij} : i, j \in V \right)$ setting
\begin{align*}
    \pi_{ij} & = 
    \begin{cases}
    \frac{q_{ij}}{-q_{ii}} & j \neq i \And q_{ii} \neq 0\\
    0 & j \neq i \And q_{ii} = 0\\
    \end{cases} \\
    \pi_{ii} & = 
    \begin{cases}
    0 & q_{ii} \neq 0 \\
    1 & q_{ii} = 0.
    \end{cases}
\end{align*}
Observe that the jump matrix corresponding to our $\hat{H}(\hat{W})$ is $\hat{K}(\hat{W})^{-1}\hat{W}$---$q_{ij} = W_{ij}$ and $-q_{ii} = k_1$---with ones on the diagonal for isolated nodes, if any.
Finally, by~\cite[Thm. 2.8.2]{Norris1997} we have that the master equation~\eqref{eq:master} defines a continuous-time Markov chain with generator $-\hat{H}(\hat{W})$ on $V$, i.e., a right-continuous process with (independent) exponential holding times of rates $k_i$.
In network terms this is an edge-centric continuous-time random walk with rate $k_i$ of leaving the $i-$th node~\cite{Masuda2017} and $\langle x_j |e^{-\tau \hat{H}(\hat{W}) } | x_i\rangle =: p_{\tau}(i, j)$ represents the transition probability from $i$ to $j$ in time $\tau$ \gb{or equivalently, for fixed initial node $i$, $e^{-\tau \hat{H}(\hat{W}) } | x_i\rangle = p_{\tau}(i, \cdot) =: p_{\tau}(i)$ is a bump function on $i$ with increasing support as $\tau$ grows~\cite{Coifman2006}. 
The diffusion distance~\eqref{eq:diffu-dist} can then be seen as a distance between bump functions and, in this case, it is useful to write its eigenmode expansion using the (inverse) graph Fourier transform~\cite{Sandryhaila2013,Masuda2017}.
The Laplacian matrix $\hat{H}(\hat{W})$ is real and symmetric, for undirected networks, it can then be diagonalized $\hat{H}(\hat{W}) = \hat{Q} \Lambda \hat{Q}^*$, where $\hat{Q} = \hat{Q}(\hat{W})$ is an $N \times N$ matrix of the orthonormalized eigenvectors $|\varphi_{\ell} \rangle, \ell = 0,1,... N-1$ of $\hat{H}(\hat{W})$ such that $\langle\varphi_{\ell}|\varphi_{\ell'} \rangle = \delta_{\ell \ell'}$, i.e., $\hat{Q}\hat{Q}^*=\hat{Q}^*\hat{Q}=\hat{I}$, and $\hat{\Lambda} = \hat{\Lambda}(\hat{W})$ is the diagonal matrix of its eigenvalues $\lambda_{\ell},\ell = 0,1,... N-1$.
Consequently, $e^{-\tau \hat{H}(\hat{W})} = \hat{Q} e^{-\tau \hat{\Lambda}} \hat{Q}^*$ and 
\begin{align*}
    p_{\tau}(i, k) & = \sum_{\ell=0}^{N-1} a_{\ell}(i; \tau) \varphi_{\ell}(k) \\
    \text{ with } & \begin{cases} 
    & a_{\ell}(i; \tau) = e^{-\tau\lambda_{\ell}} a_{\ell}(i; 0)\\
    & a_{\ell}(i; 0) = \langle x_i | \varphi_{\ell}\rangle =: \varphi_{\ell}(i)
    \end{cases}
\end{align*}
Eq.~\eqref{eq:diffu-dist} can then be re-written as
\begin{align}
    D^2_{\tau} (i, j) & = \lVert p_{\tau}(i) - p_{\tau}(j) \rVert ^ 2 = \sum_{\ell} \left(a_{\ell}(i; \tau) - a_{\ell}(j; \tau) \right)^2 \nonumber \\
    & = \sum_{\ell} e^{-2 \tau \lambda_{\ell}} \left(\varphi_{\ell}(i) - \varphi_{\ell}(j) \right)^2 \label{eq:iFGT}
\end{align}
Finally, observe that if network $G$ is undirected and fully connected then the eigenvalues of $\hat{H}(\hat{W})$ are $\lambda_0 = 0$ and $\lambda_{\ell} = N$ for all $\ell = 1, \dots, N-1$. $\varphi_{0}$ is constant and can be removed from the summation \eqref{eq:iFGT} hence $D^2_{\tau} (i, j) = e^{-2 N \tau}\sum_{\ell \geq 1}  \left(\varphi_{\ell}(i) - \varphi_{\ell}(j) \right)^2$, so that distances are determined uniquely by the eigenspace of $\lambda = N$. Equivalently, this can be seen using the fact that in this case $\hat{H}(\hat{W}) = (N-1)I - \hat{W}$ and, since $I\hat{W} = \hat{W}I$, $e^{-\tau \hat{H}(\hat{W})} = e^{-\tau (N-1)} e^{\tau \hat{W}}$, where $e^{\tau \hat{W}}$ contains the information about the number of paths of increasing length between pairs of nodes, which does not depend on its extremes for fully connected networks~\cite{Estrada2008}.}

\paragraph{Average square diffusion distance (ASDD)}
Eq.~\eqref{eq:master} maps nodes in a network to a cloud of points in the diffusion space, whose dispersion can be quantified using a generalized scalar measure of variance as the trace of their (sample) covariance matrix. 
We here show that this is equivalent to computing the ASDD. 

Let us fix $\tau>0$, so we can drop it from the notation, and call $p_{ij} = \langle x_j |e^{-\tau \hat{H}(\hat{W}) } | x_i\rangle$. Eq.~\eqref{eq:ASDD} becomes 
\begin{align*}
    \mathcal{D}^2(\tau, \hat{W}) & = \frac{1}{2 N^2} \sum_{i, j = 1}^N \sum_{k=1}^N \left(p_{ik} - p_{jk}\right)^2
\end{align*}
while the trace of the sample covariance matrix $C$ corresponding to the $N$ vectors $\left\{e^{-\tau \hat{H}(\hat{W}) } | x_i\rangle,~i=1, \dots, N\right\}$ is 
\begin{align*}
    \tr{C} = \frac{1}{N}\sum_{i = 1}^N \sum_{k=1}^N \left(p_{ik} - \bar{p}_i\right)^2.
\end{align*}
The equivalence can be proved using a known property of the variance---also known as variance deformation formula~\cite{Zhang2012}\textemdash, but it is also easily proved ``by hand'' in case particular case of undirected networks, where the Laplacian matrix $\hat{H}(\hat{W})$ is symmetric.
Firstly, observe that for all $i$ $\bar{p}_i = \frac1N \sum_{k=1}^N p_{ik} = \frac1N =: \bar{p}$
Then,
\begin{align*}
    \tr{C} & = \frac{1}{N}\sum_{i = 1}^N \left(\sum_{k=1}^N p_{ik}^2 -2\bar{p} \sum_{k=1}^N p_{ik} + \sum_{k=1}^N \bar{p}^2\right)^2 \\
    & =  \frac{1}{N} \left(\sum_{i = 1}^N m^2_i - 1\right)
\end{align*}
where we called $m^2_i = \sum_{k=1}^N p_{ik}^2$ the raw second moment of the $i-$th random vector.
Similarly, 
\begin{align*}
    \mathcal{D}^2(\tau, \hat{W}) & = \frac{1}{2 N^2} \sum_{i, j = 1}^N \sum_{k=1}^N p_{ik}^2 - 2 \sum_{k=1}^N p_{ik}p_{jk} + \sum_{k=1}^N p_{jk}^2 \\
    & = \frac{1}{2 N^2} \sum_{i, j = 1}^N \left(m_i^2 - 2(P^2)_{ij} + m_j^2 \right) \\
    & = \frac{1}{2 N^2} \left(2N \sum_{i=1}^N m_i^2 - 2\sum_{i=1}^N \sum_{j=1}^N (P^2)_{ij} \right)\\
    & =  \frac{1}{N} \left(\sum_{i=1}^N m_i^2 - 1 \right)
\end{align*}
where we used the undirected network assumption and the semigroup property of $e^{-\tau \hat{H}}$ to write $p_{jk} = p_{kj}$ yielding $\sum_{k=1}^N p_{ik}p_{jk} = (P^2)_{ij}$ and then $\sum_{j} (P^2)_{ij} = 1$.

Finally, two minor observations: firstly, here we use the biased sample covariance, but its unbiased version, with factor $\frac{1}{N-1}$ instead of $\frac1N$, can also be used provided that the sum of squared distances is also divided by $\frac{1}{N(N-1)}$ instead of $\frac{1}{N^2}$, i.e., the zeros on the diagonal are not counted in in the sum.
Secondly, the factor $\frac12$ in the ASDD definition can also be seen as a re-scaling into $[0, 1]$ of the diffusion distances, which are indeed bounded in $[0, \sqrt{2}]$~\cite{Bertagnolli2021}.

\section{Network entanglement centrality}\label{appx:entanglement}
\begin{figure}[h]
    \centering
    \includegraphics[width=0.35\textwidth]{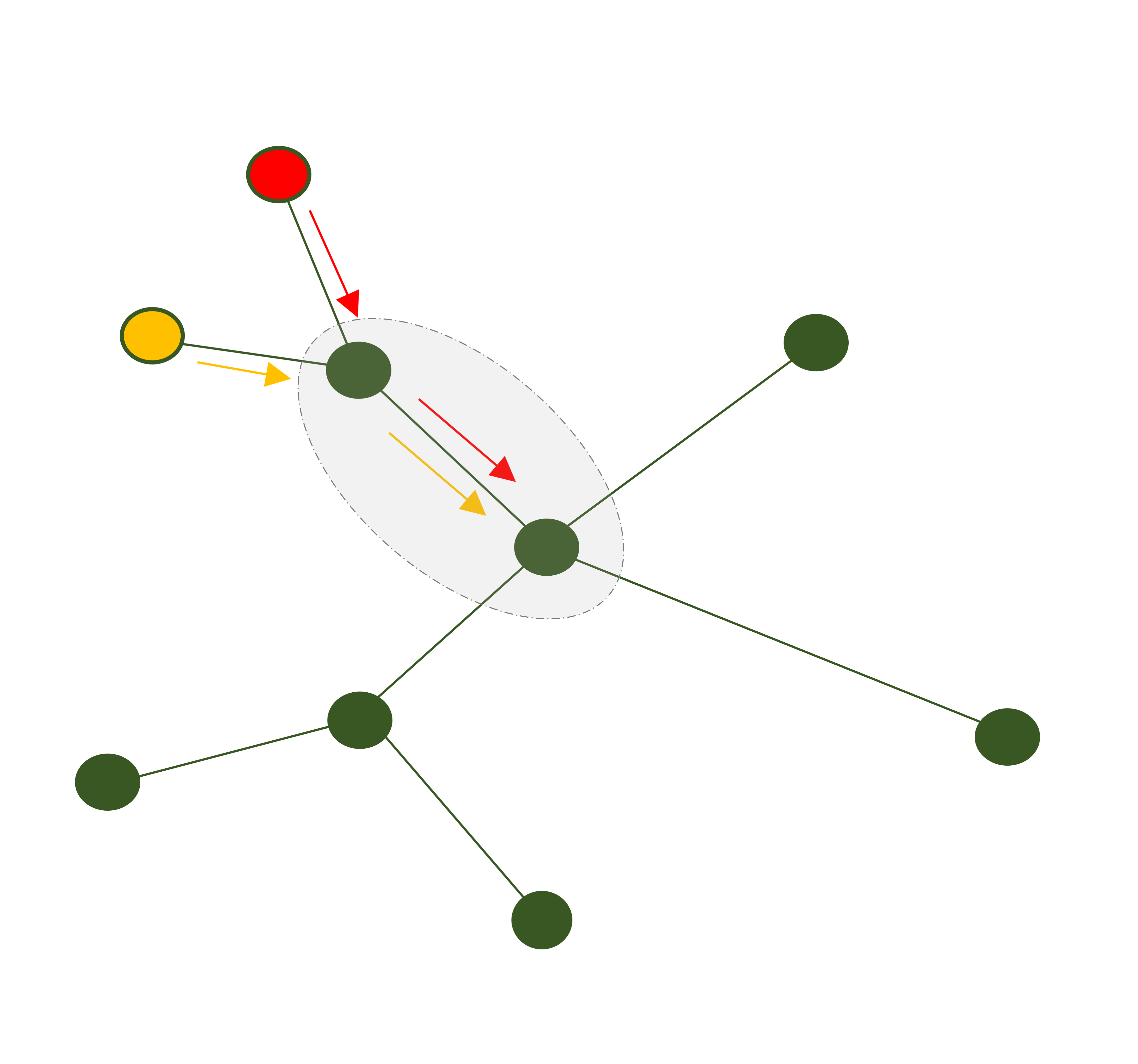}
    \caption{\textbf{Flow overlap.} Propagation of flow from two nodes, represented in red and yellow, as arrows emanating from them. The overlap region is colored in gray---i.e., the nodes within the gray zone receive information from both initiators. The overlap between the flow emanating from any pair of nodes, $i$ and $j$, can be calculated using cosine similarity between the corresponding propagation vectors $ \hat{U}(\tau,\hat{W})|x_{i}\rangle$ and $ \hat{U}(\tau,\hat{W})|x_{j}\rangle$. Conversely, average cosine dissimilarity of propagation vectors indicates the diversity of flow pathways in the system, which is proportional to the system's Von Neumann entropy~\cite{sft2020}.  }
    \label{fig:overlap}
\end{figure}

It is possible to describe the macroscopic state of complex interconnected systems, in terms of physical quantities such as entropy~\cite{de2016spectral} and free energy, in a multi-resolution approach determined by the propagation time-scale $\tau$ of signals, introduced in the text. The density matrix derived from the formalism~\cite{sft2020} reads
\begin{eqnarray}\label{eq:density}
\hat{\rho}(\tau,\hat{W})=\frac{\hat{U}(\tau,\hat{W})}{\tr{\hat{U}(\tau,\hat{W})}},
\end{eqnarray}
and the corresponding Von Neumann entropy is given by
\begin{eqnarray}
\mathcal{S}(\tau,\hat{W}) = -\tr{\hat{\rho}(\tau,\hat{W})\log_{2}{\hat{\rho}(\tau,\hat{W})}}.
\end{eqnarray}
Interestingly, the Von Neumann entropy measures the mixedness of the ensemble of stream operators directing the information propagation. In other words, it provides a proxy for diversity of flow pathways in the system, characterizing the functional diversity of nodes in sending or receiving information~\cite{sft2020}. More specifically, entropy is shown to be inversely proportional to the average overlap of the flow emanating from every pair of nodes $i$ and $j$, given by $\hat{U}(\tau,\hat{W})|x_{i}\rangle$ and $\hat{U}(\tau,\hat{W})|x_{j}\rangle$ (See Fig.~\ref{fig:overlap}). 
For instance, in a totally dismantled network of $N$ isolated nodes and no connections, the flow vectors initiated by nodes have zero overlap, leading to highest possible entropy $\log_{2}{N}$~\cite{de2016spectral} \gb{and largest possible pairwise diffusion distances $D_{\tau}(i, j) = \sqrt{2}$ for all $i, j$ and $\tau$ and, consequently, to the maximum spreading of the nodes in the diffusion space, i.e., $\mathcal{D}^2(\tau, \hat(W))=1$}. 
Conversely, the propagation vectors have the highest possible overlap in a fully connected network. \gb{Also in this case the diffusion distance is constant for all pairs of nodes and is uniquely determined by $\tau$ and by the number of nodes (see the Appendix \ref{appx:diff-geom}).} 

\begin{figure}[!hb]
    \centering
    \includegraphics[width=0.5\textwidth]{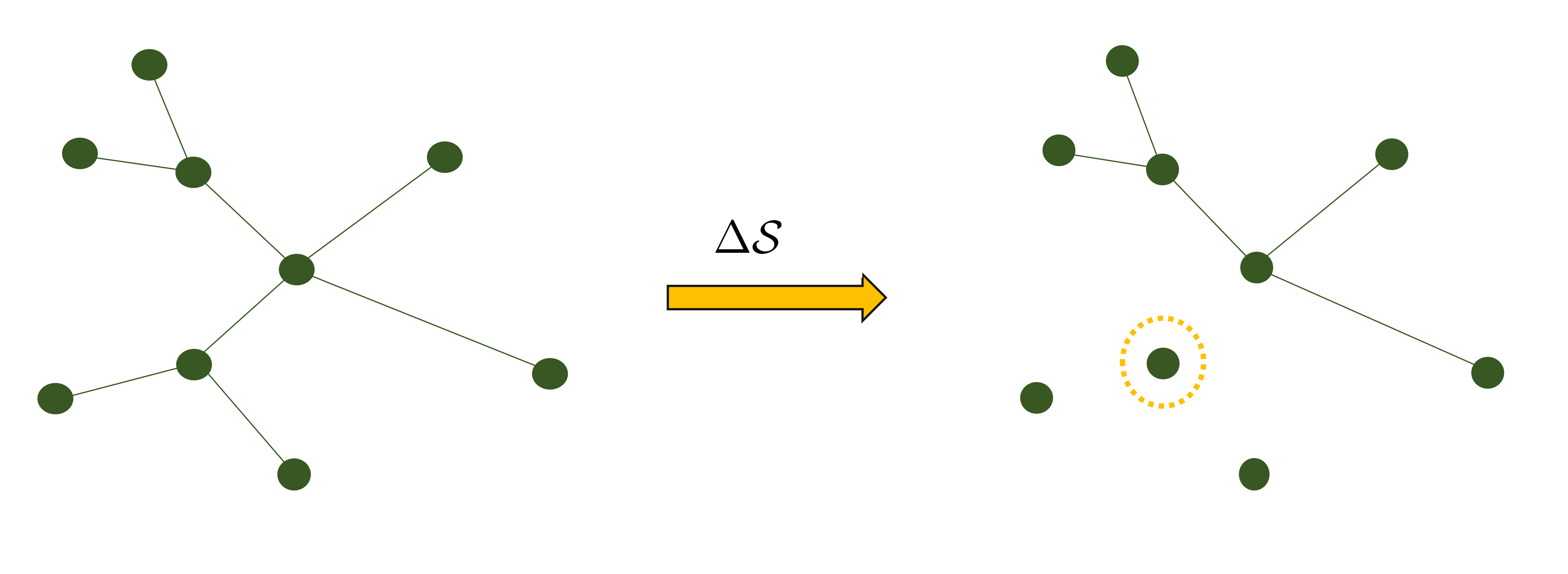}
    \caption{\textbf{Detachment process.} Node-network entanglement is defined in terms of the entropy change due to detachment of one node. Network Von Neumann entropy provides a proxy for the diversity of flow pathways~\cite{sft2020}: Therefore, the node having highest entanglement with the network is important, as its removal severely affects information exchange. This makes entanglement a powerful measure to identify the nodes that are functionally critical for information dynamics and to find attack strategies to functionally dismantle networks. }
    \label{fig:detachment}
\end{figure}

Assume the initial entropy a network, before damage, is $\mathcal{S}^{(0)}(\tau,\hat{W})$. At the final stage, where all the nodes are detached---i.e., all the links connecting them are removed---and the full structural and functional dismantling is achieved, it follows $\mathcal{S}^{(f)}(\tau,\hat{W})=\log_{2}{N}$. If we want the fastest transition from $\mathcal{S}^{(0)}(\tau,\hat{W})$ to $\mathcal{S}^{(f)}(\tau,\hat{W})$, in what order should we rank and, respectively, disrupt the units? The simplest answer to this question is that at each step, we remove the node whose removal has the maximum increase in the Von Neumann entropy. For instance, let $\mathcal{S}^{(m-1)}(\tau,\hat{W})$ be the entropy of the network before the removal of $m$-th node and $\mathcal{S}^{(m)}_{i}(\tau,\hat{W})$ be the entropy of the network if one removes the $i$-th node at the $m$-th step (See Fig.~\ref{fig:detachment}). The entanglement~\cite{entanglement2021} between the $i$-th node and the network at that step is defined as

\begin{equation}\label{eq:entanglement}
\epsilon_{i}^{(m-1)}(\tau,\hat{W}) = \mathcal{S}^{(m)}_{i}(\tau,\hat{W}) - \mathcal{S}^{(m-1)}(\tau,\hat{W}),
\end{equation}
that can be used as a measure quantifying the importance of that node for diversity of flow pathways. At each step, we identify the node with maximum entanglement by
\begin{eqnarray}
\max_{i}[\epsilon_{i}^{(m-1)}(\tau,\hat{W})] = \max_{i}[\mathcal{S}^{(m)}_{i}(\tau,\hat{W})].
\end{eqnarray}

Note that the second term in~\eqref{eq:entanglement} is independent of the node to be removed and, therefore, it vanishes from the maximization. Assuming that the $i$-th node has the highest entanglement at the $m$-th step, the entropy at $m$-th step reads $\mathcal{S}^{(m)}(\tau,\hat{W}) =\mathcal{S}^{(m)}_{i}(\tau,\hat{W}) $. Note that the algorithm must exclude the nodes that are already isolated, before performing the maximization, since the detachment process does not make sense for them. 

It is worth mentioning that the detachment process considered here (See Fig.~\ref{fig:detachment}) is slightly different from the original definition~\cite{entanglement2021}, where the node was removed with its incident edges shaping an independent star network. In fact, the previous definition has a nice property, used to prove that the behavior of entanglement of a node at very small $\tau$ is determined by its degree. However, it seems also natural to think that the detached node becomes isolated from the rest of the network, as considered here.  

\onecolumngrid
\section{Definitions}
\rev{
In this appendix we collect some common expressions used in the literature and in this work, with the relative (tentative) definitions and examples from different fields.

\begin{longtable}{p{2.5cm} p{13cm}}
\textbf{Expression} & \textbf{Definition and examples} \\
\endhead
\hline
\multirow{2}{*}{Structure} &
  The arrangement of and relations between the components of complex systems, often modeled in terms of networks where components are nodes and their connections are links. Sometimes topological structure, or simply "the network topology", is used to indicate the arrangement of dyadic interactions between units, in contrast with "weighted structure", which also includes the intensity of the interactions.\\ 
  &
  \begin{itemize}
    \item In physics: The connections between states of physical systems determining the rates of transitions between them.
    \item In biology: The network encoding connections between cells, organs or species.
    \item In chemistry: The the interrelations between chemical compounds in a chemical reaction network.
    \item In social sciences: The relationships between individuals in a social network.
    \item In transportation systems: The network of transportation routes connecting districts, urban areas, regions, countries or continents.
\end{itemize} \\ \hline
\multirow{2}{2.5cm}{Dynamical process} &
  Quantities or fields change with respect to time, according to rules imposed by differential equations each known as a dynamical process. \\
 &
  \begin{itemize}
    \item In physics: The thermalization protocol describing the transitions between physical states of a system.
    \item In biology: The biochemical equation governing the spreading of chemicals or electrical signals among cells, organs or species.
    \item In chemistry: The reaction-diffusion equations describing the behavior of the population corresponding to a chemical reaction network.
    \item In social sciences: The consensus dynamics between individuals in a social network, or the dynamical equations describing the spread of pathogens or news between them.
    \item In transportation systems: The equation describing the flow of people or goods through the network of transportation routes connecting districts, urban areas, regions, countries or continents.
\end{itemize} \\ \hline
\multirow{2}{2.5cm}{Information exchange} & A general term to describe the effect of components of complex systems on each other, often modeled in terms of the flow of a physical quantity between pairs of nodes, governed by dynamical processes. \textit{Communication} is here used as a synonym.\\
& \begin{itemize}
    \item In physics: The exchange of particles between two physical states, induced by a themalization protocol or external forces.
    \item In biology: The exchange of electrochemical signals between two cells, organs or species.
    \item In chemistry: The influence of two chemical compounds on each other, leading to a change in their populations, in a chemical reaction network.
    \item In social sciences: The consensus dynamics between individuals in a social network, or the dynamical equations how one individual infects the other with a pathogen or informs the other about a news.
    \item In transportation systems: The exchange of people or goods between two nodes, through the network of transportation routes connecting districts, urban areas, regions, countries or continents.
\end{itemize} \\ \hline
\multirow{2}{2.5cm}{Information flow} & Emanation of a quantity or field, whose exchange between the components proxies their communications, from a source, often considered to be one of the components, into the system, through the links. \\
& \begin{itemize}
    \item In physics: The flow of particles from one physical state into others, induced by a themalization protocol or external forces.
    \item In biology: The flow of electrochemical signals from cells, organs or species.
    \item In chemistry: The impact of a chemical compound on the populations of others, in a chemical reaction network.
    \item In social sciences: The flow of pathogen or news from an individual to the rest.
    \item In transportation systems: The flow of people or goods from one of the nodes, through the network of transportation routes connecting districts, urban areas, regions, countries or continents.
\end{itemize} \\ \hline
\multirow{2}{2.5cm}{Impairment of information flow} & When damage significantly lowers the exchange of the field, that proxies communications, between the components across the system.\\
& 
\begin{itemize}
    \item In physics: When external perturbations lower the flow of particles from one physical state into others, induced by a themalization protocol or external forces.
    \item In biology: When damage lowers the flow of electrochemical signals from cells, organs or species.
    \item In chemistry: When external perturbations lowers the impact of a chemical compound on the populations of others, in a chemical reaction network.
    \item In social sciences: When damage lowers the flow of pathogen or news from an individual to the rest.
    \item In transportation systems: When damage lowers the flow of people or goods from one of the nodes, through the network of transportation routes connecting districts, urban areas, regions, countries or continents.
\end{itemize} \\ \hline
\multirow{2}{*}{Function} & System specific tasks expected to be performed, that can involve a single component or a group of components that exchange information with each other.\\
& 
\begin{itemize}
    \item In physics: The activity of a classical or quantum heat engine with certain properties like power and efficiency.
    \item In biology: The physiological activity of a cell, organ, system or body.
    \item In chemistry: The characteristic behavior of a chemical compound or groups of chemicals linked in a chemical reaction network.
    \item In social sciences: The activity or behavior of an individual or a group of them in a society.
    \item In transportation systems: Financial activities that depend on the flow of people or goods from one of the nodes, through the network of transportation routes connecting districts, urban areas, regions, countries or continents.
\end{itemize} \\ \hline
\multirow{2}{*}{Operation} & While function and operation are used interchangeably in the literature, the more precise definition of operation is the method and mechanism by which a system or a part of it performs its function.\\ \\ \hline
\multirow{2}{2.5cm}{Degradation of function} & When damage disturbs the components performing a function or impairs the flow of information between groups of component, preventing them from performing a function.\\
& 
\begin{itemize}
    \item In physics: Perturbed activity of a classical or quantum heat engine that reflects in certain properties like power and efficiency.
    \item In biology: Significant perturbation of physiological activity of a cell, organ or body.
    \item In chemistry: Significant perturbation of the dynamics of a groups of chemicals linked in a chemical reaction network.
    \item In social sciences: Significant disturbance in activity of an individual or a group of them in a society.
    \item In transportation systems: Hindering of the flow of people or goods from one of the nodes, through the network of transportation routes connecting districts, urban areas, regions, countries or continents.
\end{itemize}
\\ \hline
\end{longtable}
}

\end{document}